\documentclass[prx,nofootinbib,twocolumn,showkeys,superscriptaddress,preprintnumbers,floatfix]{revtex4-1}

\usepackage{dynlearn}
\fxsetup{status=draft}
\usepackage{empheq}

\usepackage{kbordermatrix}

\newcommand{\Shannon}{\H}
\newcommand{\kB}{k_\text{B}}
\newcommand{\Wdiss}{W_\text{diss}}

\newcommand{\St}{\mathcal{S}}
\newcommand{\MSt}{\mathcal{M}}
\newcommand{\SSet}{\boldsymbol{\mathcal{S}}}

\newcommand{\MSet}{\boldsymbol{\mathcal{M}}}

\newcommand{\actual}[1][0]{\boldsymbol{\mu}_{#1}}

\newcommand{\Reverse}{\protect\reflectbox{$\mathbf{R}$}}

\newcommand{\IverL}{\bigl[ \!\! \bigl[}
\newcommand{\IverR}{\bigr] \!\! \bigr]}

\newcommand{\Left}{\text{L}}    \newcommand{\Right}{\text{R}}

\renewcommand{\H}{\operatorname{H}}

\newcommand{\st}{\causalstate}

\let\st\relax
\newcommand{\st}{s} 

  \newcommand{\protocol}{x}

\newcommand{\HamSusLab}{\mathcal{H}_{SL}}

\newcommand{\Traj}{\overrightarrow \St} \newcommand{\traj}{\overrightarrow \st}

\newcommand{\Ent}{H}
\newcommand{\Fneq}{\mathcal{F}}
\newcommand{\work}{W}
\newcommand{\workavg}{\langle W \rangle}
\newcommand{\entprod}{\beta W_\text{diss}}
\newcommand{\entprodavg}{\beta \langle W_\text{diss} \rangle}

\newcommand{\forinistdist}{\actual[0]} \newcommand{\revinistdist}{\actual[\tau]} 

\newcommand{\forP}{P}
\newcommand{\revP}{R}

\newcommand{\mst}{m}
\newcommand{\msttwo}{m'}

\newcommand{\forPcoarse}{P'}

\newcommand{\forinistdistcoarse}{\actual[0]'}

\newcommand{\forPcoarseto}[2]{\forPcoarse({#1} \to {#2})}
\newcommand{\comp}{C}
\newcommand{\error}{\epsilon}
\newcommand{\errorto}[2]{\epsilon_{{#1}\to{#2}}}
\newcommand{\erroravg}{\langle \error \rangle}

\newcommand{\tspworkbound}{\langle W \rangle_\text{min}^\text{t-sym}}
\newcommand{\tspworkboundapprox}{\langle W \rangle_\text{min}^\text{approx}}

\usepackage{setspace}

\begin{document}

\def\ourTitle{Refining Landauer's Stack: \\[0.06in]
Balancing Error and Dissipation When Erasing Information
}

\def\ourAbstract{Nonequilibrium information thermodynamics determines the minimum energy
dissipation to reliably erase memory under time-symmetric control protocols. We
demonstrate that its bounds are tight and so show that the costs overwhelm
those implied by Landauer's energy bound on information erasure. Moreover, in
the limit of perfect computation, the costs diverge. The conclusion is that
time-asymmetric protocols should be developed for efficient, accurate
thermodynamic computing. And, that Landauer's Stack---the full suite of
theoretically-predicted thermodynamic costs---is ready for experimental test
and calibration.
}

\def\ourKeywords{nonequilibrium steady state, thermodynamics, dissipation, entropy
  production, Landauer bound
}

\hypersetup{
  pdfauthor={James P. Crutchfield},
  pdftitle={\ourTitle},
  pdfsubject={\ourAbstract},
  pdfkeywords={\ourKeywords},
  pdfproducer={},
  pdfcreator={}
}

\title{\ourTitle}

\author{Gregory W. Wimsatt}
\email{gwwimsatt@ucdavis.edu}
\affiliation{Complexity Sciences Center and Physics Department,
University of California at Davis, One Shields Avenue, Davis, CA 95616}

\author{Alexander B. Boyd}
\email{abboyd@ucdavis.edu}
\affiliation{Complexity Institute and School of Physical and Mathematical Sciences,
Nanyang Technological University,
637371 Singapore, Singapore}

\author{Paul M. Riechers}
\email{pmriechers@gmail.com}
\affiliation{Complexity Institute and School of Physical and Mathematical Sciences,
Nanyang Technological University,
637371 Singapore, Singapore}

\author{James P. Crutchfield}
\email{chaos@ucdavis.edu}
\affiliation{Complexity Sciences Center and Physics Department,
University of California at Davis, One Shields Avenue, Davis, CA 95616}

\date{\today}
\bibliographystyle{unsrt}

\begin{abstract}
\ourAbstract
\end{abstract}

\keywords{\ourKeywords}

\preprint{\arxiv{2011.XXXXX}}

\date{\today}
\maketitle

\setstretch{1.1}

\section{Introduction}

In 1961, Landauer identified a fundamental energetic requirement to perform
logically-irreversible computations on nonvolatile memory \cite{Land61a}.
Focusing on arguably the simplest case---erasing a bit of information---he
found that one must supply at least $\kB T \ln 2$ work energy ($\approx
10^{-21} J$ at room temperature), eventually expelling this as heat. (Here,
$\kB$ is Boltzmann's constant and $T$ is the temperature of the computation's
ambient environment.)

Notably, though still underappreciated, Landauer had identified a
thermodynamically-reversible transformation. And so, no entropy actually need
be produced---energy is not irrevocably dissipated---at least in the
quasistatic, thermodynamically-reversible limit required to meet Landauer's
bound.

Landauer's original argument appealed to equilibrium statistical mechanics.
Since his time, advances in nonequilibrium thermodynamics, though, showed that
his bound on the required work follows from a modern version of the
Second Law of thermodynamics \cite{Parr15a}. (And, when the physical
substrate's dynamics are taken into account, this is the \emph{information
processing Second Law} (IPSL) \cite{Boyd15a}.) These modern laws clarified many
connections between information processing and thermodynamics, such as
dissipation bounds due to system-state coarse-grainings \cite{Gome08a},
nanoscale information-heat engines \cite{Deff2013}, the relation of dissipation
and fluctuating currents \cite{Li19}, and memory design \cite{Stil20}.

Additional scalings recently emerged between computation time, space,
reliability, thermodynamic efficiency, and robustness of information storage
\cite{Gopa15a,Lahi18a,Boyd18a}. In contrast to Landauer's bound, these
tradeoffs involve thermodynamically-irreversible processes, implying that
entropy production and therefore true heat dissipation is generally required
depending on either practicality or design goals.

In addition to these tradeoffs, it is now clear that substantial energetic
costs are incurred when using logic gates and allied information-processing
modules to construct a computer. Especially so, when compared to custom
designing hardware to optimally implement a particular computation
\cite{Boyd17a}.

\begin{figure*}[t]
\centering
\includegraphics[width=0.8\linewidth]{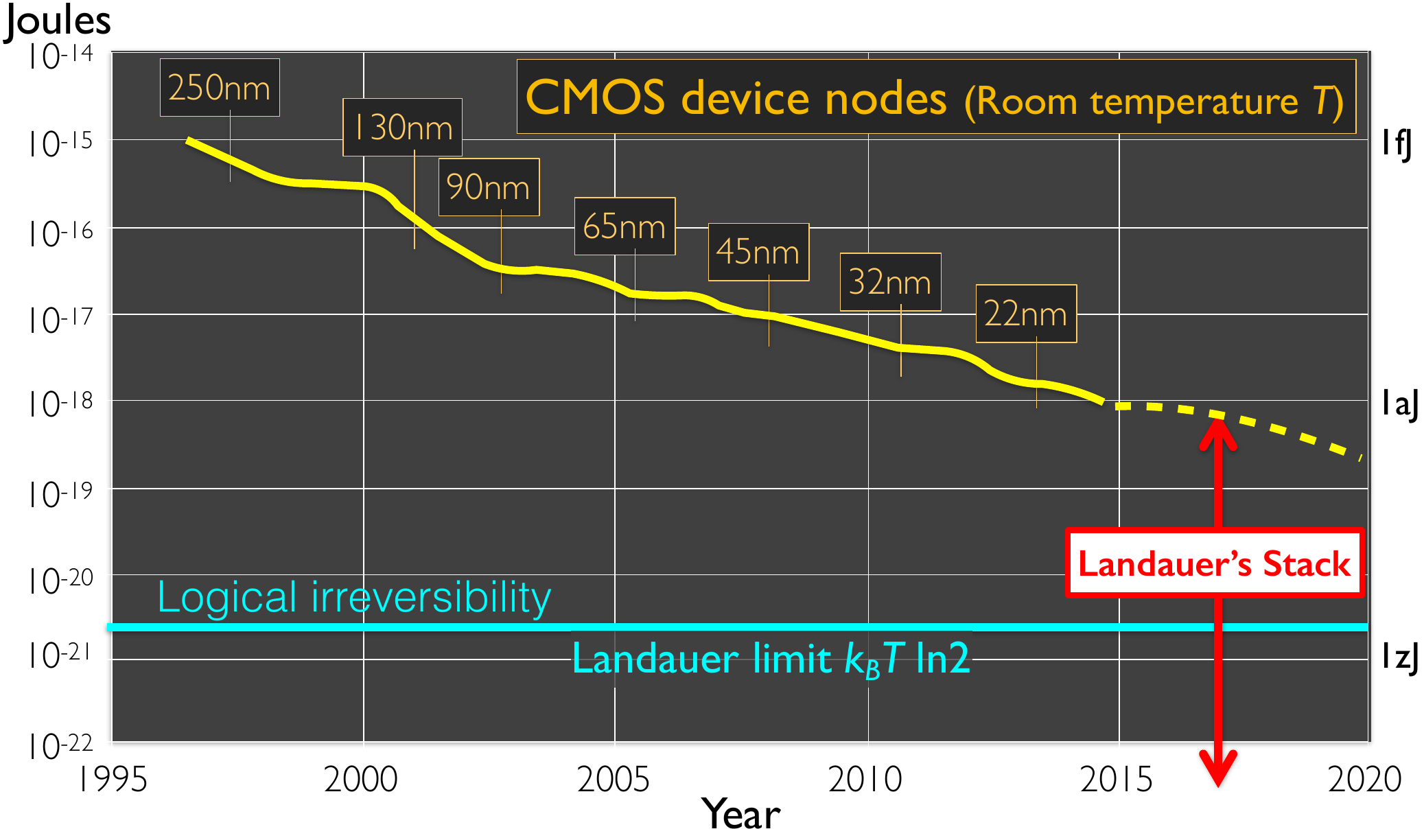}
\caption{Historical trends in thermodynamic costs of performing elementary
	logic operations in CMOS technology quoted in energy dissipated (joules) per
	logic operation. Contemporary experimentally-accessible thermal resolution
	is approximately $10^{-24}$ joules.
	Landauer's Stack---Thermodynamic hierarchy of predicted ``overhead'' energy
	expenditures \emph{due to information processing} that underlie
	contemporary digital computing including Landauer's Principle of logical
	irreversibility \cite{Land61a}, now seen as a consequence of the broader
	information processing Second Law $\langle W \rangle \leq \kB T \ln \Delta
	\hmu$ \cite{Boyd15a}:
	(a) Nonreciprocity \cite{Riec19b};
	(b) Computation rate $1/\tau$ \cite{Lahi18a,Boyd18a};
	(c) Accuracy: $- \ln \epsilon$ \cite{Riec19b};
	(d) Storage stability;
	(e) Circuit modularity \cite{Boyd17a};
	(f) Transitions between nonequilibrium steady-state storage states
	\cite{Boyd16e,Riec16b}; and
(g) Quantum coherence \cite{Loom19b}.
	(2015 and prior portion of figure courtesy M. L. Roukes, data compiled from
	\cite[and citations therein]{ITRS15a}. Landauer's Stack cf. Table I in Ref.
	\cite{Cont99a}.)
	CMOS technology change to 3D device nodes around 2015 make linear feature
	size and its relation to energy costs largely incomparable afterwards
	\cite{ITRS20a,ITRS20b,ITRS20c,Shal20a}.
	There are, of course, other sources of energy dissipation in CMOS such as
	leakage currents that arise when electrons tunnel from gate to drain
	through a thin gate-oxide dielectric. Thermodynamically, this source is
	kind of ``housekeeping heat'', necessary to support the substrate's
	electronic properties but not directly due to information processing.
	}
\label{fig:Stack} 
\end{figure*}

Taken altogether these costs constitute a veritable \emph{Landauer's Stack} of
the information-energy requirements for thermodynamic computing. Figure
\ref{fig:Stack} illustrates Landauer's Stack in the light of historical trends
in the thermodynamic costs of performing elementary logic operations in CMOS
technology. The units there are joules dissipated per logic operation. We take
Landauer's Stack to be the overhead including Landauer's bound ($\kB T \ln 2$
joules) up to the current (year 2020) energy dissipations \emph{due to
information processing}. Thus, the Stack is a hierarchy of energy expenditures
that underlie contemporary digital computing---an arena of
theoretically-predicted and as-yet unknown thermodynamic phenomena waiting
detailed experimental exploration.

To account for spontaneous deviations that arise in small-scale systems, the
Second Laws are now most properly expressed by exact equalities on probability
distributions of possible energy fluctuations. These are the \emph{fluctuation
theorems} \cite{Klag13a}, from which the original Laws (in fact, inequalities)
can be readily recovered. Augmenting the Stack, fluctuation theorems apply
directly to information processing, elucidating further thermodynamic
restrictions and structure \cite{Saga12a, Beru12a, Engl15, Sair19c}.

The result is a rather more complete accounting for the energetic costs of
thermodynamic computation, captured in the refined Landauer's Stack of Fig.
\ref{fig:Stack}. In this spirit, here we report new bounds on the work required
to compute in the very important case of computations driven externally by
time-symmetric control protocols \cite{Riec19b}. In surprising contrast to the
fixed energy cost of erasure identified by Landauer, here we demonstrate that
the scaling of the minimum required energy \emph{diverges as a function of
accuracy} and so can dominate Landauer's Stack. This serves the main goal in
the following to validate and demonstrate the tightness of Ref.
\cite{Riec19b}'s thermodynamic bounds and do so in Landauer's original setting
of information erasure.

In essence, our argument is as follows. Energy dissipation in thermodynamic
transformations is strongly related to entropy production. The fluctuation
theorems establish that entropy production depends on both forward and reverse
dynamics. Thus, when determining bounds on dissipation in thermodynamic
computing, one has to examine both when the control protocol is applied in
forward and reverse. By considering time-symmetric protocols we substantially
augment Landauer and Bennett's dissipation bound on logical irreversibility
\cite{Benn82} with dissipation due to logical nonselfinvertibility
(aka nonreciprocity).

Why time-symmetric protocols? Modern digital computers are driven by sinusoidal
line voltages and square-wave clock pulses---time-symmetric control signals.
And so, modern digital computers obey the Ref. \cite{Riec19b}'s
error-dissipation trade-off. Moreover, the costs apply to even the most basic
of computational tasks---such as bit erasure. Here, we present protocols for
time-symmetrically implementing erasure in two different frameworks and
demonstrate that both satisfy the new bounds.  Moreover, many protocols
approach the bounds quite closely, indicating that they may in be fact be
broadly achievable.

After a brief review of the general theory, we begin with an analysis of
erasure implemented with the simple framework of two-state rate equations,
demonstrating the validity of the bound for different protocols of increasing
reliability. We then expand our framework to fully simulated collections of
particles erased in an underdamped Langevin double-well potential, seeing the
same faithfulness to the bound for a wide variety of different erasure
protocols. We conclude with a call for follow-on efforts to analyze even more
efficient computing that can arise from \emph{time-asymmetric} protocols.

\section{Dissipation In Thermodynamic Computing}

Consider a universe consisting of a computing device---the \emph{system under
study} (SUS), a \emph{thermal environment} at fixed inverse temperature
$\beta$, and a \emph{laboratory device} (lab) that includes a \emph{work
reservoir}. The set of possible microstates for the SUS is denoted $\SSet$,
with $\st$ denoting an individual SUS microstate. The SUS is driven by
a \emph{control parameter} $\protocol$ generated by the lab. The SUS is also in
contact with the thermal environment.

The overall evolution occurs from time $t = 0$ to $t = \tau$ and is
determined by two components. The first is the SUS's Hamiltonian
$\HamSusLab(\st, \protocol)$ that specifies its interaction with the lab
device and determines (part of) the rates of change of the SUS coordinates
consistent with Hamiltonian mechanics. We refer to the possible values of the
Hamiltonian as the \emph{SUS energies}. The second component is the thermal
environment which exerts a stochastic influence on the system dynamics.

We prepare the lab to guarantee that a specific control parameter value $x(t)$
is applied to the SUS at every time $t$ over the time interval $t \in (0,
\tau)$. That is, the control parameter evolves deterministically as a function of
time. The deterministic trajectory taken by the control parameter $x(t)$ over
the computation interval is the \emph{control protocol}, denoted by
$\overrightarrow{x}$. The SUS microstate $s(t)$ exhibits a response to the
control protocol, over the interval following a stochastic trajectory denoted
$\overrightarrow{s}$.

For a given microstate trajectory $\traj $, the net energy transferred from the
lab to the SUS is defined as the \emph{work}, which has the following form
\cite{Deff2013}:
\begin{align*}
\work(\traj) = \int_0^\tau dt\, \dot \protocol (t)
  \frac{\partial\HamSusLab}{\partial \protocol}\Big|_{(s(t), \protocol(t))}
  ~.
\end{align*}
This is the energy accumulated in the SUS directly caused by changes in the the control parameter.

Given an initial microstate $\st_0$, the probability of a microstate trajectory
$\traj$ conditioned on starting in $\st_0$ is denoted:
\begin{align*}
\forP(\traj | \st_0) = \Pr_{ \overrightarrow{x}}(\Traj=\traj | \St_0=\st_0)
  ~.
\end{align*}
With the SUS initialized in microstate distribution $\forinistdist$, the
unconditioned \emph{forward process} gives the probability of trajectory
$\traj$:
\begin{align*}
\forP(\traj)
= \forP(\traj | s(0)) \forinistdist(s(0))
  ~.
\end{align*}

\emph{Detailed fluctuation theorems} (DFTs) \cite{Jarz00a, Croo99a} determine
thermodynamic properties of the computation by comparing the forward process to
the \emph{reverse process}. This requires determining the conditional
probability of trajectories under time-reversed control:
\begin{align*}
R(\traj | \st_0) = \Pr_{\Reverse \overrightarrow{x}}(\Traj=\traj | \St_0=\st_0)
  ~.
\end{align*}
The reverse control protocol is $\Reverse x(t)=x(\tau-t)^\dagger$, where
$x^\dagger$ is $x$, but with all time-odd components (e.g., magnetic field)
flipped in sign. And, the \emph{reverse process} results from the application
of this dynamic to the final distribution $\revinistdist$ of the forward
process with microstates conjugated:
\begin{align*}
\revP(\traj) &= R(\traj | s(0)) \revinistdist(s(0)^\dagger)
~.
\end{align*}
The Crooks DFT \cite{Croo99a} then gives an equality on
both the dissipated work (or entropy production) that is produced as well as
the required work for a given trajectory induced by the protocol:
\begin{align*}
	\entprod(\traj)
&= \ln \frac{\forP(\traj)}{\revP(\Reverse \traj)} .
\end{align*}
$\Reverse \traj$ here is itself a SUS microstate trajectory with $\Reverse
s(t) = s(\tau-t)^\dagger$.

Due to their practical relevance, we consider protocols that are symmetric
under time reversal $\Reverse \overrightarrow{x}=\overrightarrow{x}$. That is,
the reverse-process probability of trajectory $\traj$ conditioned on starting
in microstate $\st_0$ is the same as that of the forward process:
\begin{align*}
\revP(\traj | \st_0) = \forP(\traj | \st_0)
	~.
\end{align*}
However, the unconditional reverse process probability of the trajectory
$\traj$ is then:
\begin{align*}
\revP(\traj) &= \forP(\traj | s(0)) \revinistdist(s(0)^\dagger)
~.
\end{align*}
This leads to a version of Crook's DFT that can be used to set modified bounds
on a computation's dissipation:
\begin{align*}
	\entprod(\traj) = \ln \frac{\forP(\traj | s(0)) \actual(s(0))}
	{\forP(\Reverse \traj | s(\tau)^\dagger) \actual[\tau](s(\tau))}
	~.
\end{align*}

Suppose, now, that the final and initial SUS Hamiltonian configurations
$\HamSusLab(s,x(\tau))$ and $\HamSusLab(s,x(0))$ are both designed to store the
same information about the SUS. The SUS microstates are partitioned into
locally-stable regions that are separated by large energy barriers in these
energy landscapes. On some time scale, a state initialized in one of these
regions has a very low probability of escape and instead locally equilibrates
to its locally-stable region. These regions can thus be used to store
information for periods of time controlled by the energy barrier heights.
Collectively, we refer to these regions as \emph{memory states} $\MSet$.

Then the probability of the system evolving to a memory state $\msttwo \in
\MSet$ given that it starts in a memory state $\mst \in \MSet$ under either the
forward or reverse process is:
\begin{align*}
	\forPcoarse(\mst \rightarrow \msttwo)
	&= \frac{\int d\traj \IverL s(0) \in \mst \land s(\tau) \in \msttwo \IverR \forP(\traj)}{\int d\traj \IverL s(0) \in \mst \IverR \forP(\traj)}
	~,
\end{align*}
where $\IverL E \IverR$ evaluates to one if expression $E$ is true and zero
otherwise.

Reference \cite{Riec19b}'s Eq. (15) bounds the average entropy production
derived under weaker restrictions than assumed here. Applying it and
simplifying, we obtain:
\begin{align}
\entprodavg
  & \geq \Delta H(\MSt(t))
  + \!\! \sum_{\mst \in \MSet} \!\! \forinistdistcoarse(\mst)
	\!\! \sum_{\msttwo \in \MSet} \!\! d(\mst, \msttwo)
	,
\label{eq:TSP_entprod_bound}
\end{align}
where:
\begin{align*}
\forinistdistcoarse(\mst)
  & = \int d\st \IverL \st \in \mst \IverR \forinistdist(\st)
  ~, \\
d(\mst, \msttwo) & =
  \forPcoarse(\mst \rightarrow \msttwo)
  \ln \frac{\forPcoarse(\mst \rightarrow \msttwo)}
  {\forPcoarse(\msttwo \rightarrow \mst)}
  ~,
\end{align*}
assuming time-reversal invariant memories,
and
$\Delta \Ent(\MSt(t))$ is the change in Shannon entropy of the memory
distribution.

To simplify the development, suppose that the energy landscape of each memory
state looks the same locally. That is, up to translation and possibly
reflection and rotation, each memory state spans the same range in microstate
space and has the same energies at each of those states. Further, suppose that
the SUS starts and ends in a metastable equilibrium distribution, differing
from global equilibrium only in the weight that each memory state is given in
the distribution. Otherwise the distribution looks identical to the global
equilibrium at the local scale of any memory state. This ensures that the
average change in SUS energy is zero, simplifying the average nonequilibrium
free energy \cite{Riec19b}:
\begin{align*}
\Delta \Fneq = -\beta^{-1} \Delta \Ent(\MSt_t)
  ~.
\end{align*}

Recalling that $\beta \braket{\Wdiss (\traj)} = \beta ( \braket{W (\traj)} -
\Delta \mathcal{F} )$ and appealing to the inequality in Eq.
(\ref{eq:TSP_entprod_bound}), we find a simple bound on the average work over
the protocol:
\begin{align}
\label{eq:TSP_work_bound}
\beta \workavg & \geq
  \sum_{\mst \in \MSet} \forinistdistcoarse(\mst)
  \sum_{\msttwo \in \MSet} d(\mst, \msttwo) \\
  & \equiv \beta \tspworkbound
  \nonumber
  ~.
\end{align}
This provides a bound on the work that depends solely on the logical operation
of the computation, but goes beyond Landauer's bound.

Since we are addressing modern computing, we consider processes that
approximate deterministic computations on the memory states. For such
computations there exists a computation function $\comp: \MSet \to \MSet$ such
that the physically-implemented stochastic map approximates the desired
function up to some small error. That is, $\forPcoarse(\mst \rightarrow
\comp(\mst)) = 1 - \error_\mst$ where $0 < \error_\mst \ll 1$. In fact, we
require all relevant errors to be bound by a small error-threshold $\error \ll
0$. That is, for all $\comp(\mst) \neq \msttwo$, let $\forPcoarse(\mst,
\msttwo) = \errorto{\mst}{\msttwo}$ so that $0 \leq \sum_{\msttwo \neq
\comp(\mst)} \errorto{\mst}{\msttwo} = \error_\mst \leq \error \ll 1$.

We can then simplify Eq. (\ref{eq:TSP_work_bound})'s bound in the limit of
small $\error$.  First, we show that $d(\mst, \msttwo) \geq 0$ for any pair of
$\mst, \msttwo$ in the small $\error$ limit, where we have:
\begin{align*}
d(\mst, \msttwo)
  & = \forPcoarseto{\mst}{\msttwo} \ln \frac{\forPcoarseto{\mst}{\msttwo}}{\forPcoarseto{\msttwo}{\mst}} \\
  & \geq \forPcoarseto{\mst}{\msttwo} \ln \forPcoarseto{\mst}{\msttwo}
  ~.
\end{align*}
If $\comp(\mst) = \msttwo$, then $\forPcoarseto{\mst}{\msttwo} = 1 -
\error_\mst \geq 1 - \error$, so that:
\begin{align*}
	d(\mst, \msttwo) &\geq (1-\error) \ln (1 - \error)
	~,
\end{align*}
which vanishes as $\error \to 0$.
And, if $\comp(\mst) \neq \msttwo$, then $\forPcoarseto{\mst}{\msttwo} =
\errorto{\mst}{\msttwo}$, so that:
\begin{align*}
	d(\mst, \msttwo)
	&\geq \errorto{\mst}{\msttwo} \ln \errorto{\mst}{\msttwo}
\end{align*}
which also vanishes as $\error \to 0$.
Setting this asymptotic lower bound on the dissipation of each transition
facilitates isolating divergent contributions, such as those we now consider.

An \emph{unreciprocated} memory transition $\comp(\mst) = \msttwo$ is one that
does not map back to itself: $\comp(\msttwo) \neq \mst$.  The contribution to
the dissipation bound is:
\begin{align*}
d(\mst, \msttwo)
	& = (1-\error_\mst) \ln \frac{1-\error_\mst}{\errorto{\msttwo}{\mst}} \\
	& \geq (1-\error) \ln \frac{1-\error}{\error}
  ~.
\end{align*}
As $\error \to 0$, this gives:
\begin{align}
d(\mst, \msttwo) \geq \ln \error^{-1}
  ~.
\label{eq:d2_bound}
\end{align}
That is, as computational accuracy increases ($\error \to 0$), $d(\mst,
\msttwo)$ diverges. This means the minimum-required work (Eq.
(\ref{eq:TSP_work_bound})) must then also diverge.

We then arrive at our simplified bound for the small-$\error$ high-accuracy
limit from Eq. (\ref{eq:TSP_work_bound})'s inequality on dissipation by only
including the contribution from unreciprocated transitions $m' = \comp(m)$ for
which $m \neq \comp(m')$:
\begin{align}
\label{eq:TSP_work_bound_approx}
\beta \workavg
  & \geq \ln (\error^{-1}) \sum_{\mst \in \MSet}
  \forinistdistcoarse(\mst)
  \IverL \comp(\comp(\mst)) \neq \mst \IverR \\
  & \equiv \beta \tspworkboundapprox
  ~.
  \nonumber
\end{align}
In this way, we see how computational accuracy drives a thermodynamic cost that
diverges, overwhelming the Landauer-erasure cost.

To be quantitative beyond a formal divergence, consider contemporary DRAM memory
which exhibits a range of ``soft'' error rates around $10^{-22}$ failures per
write operation \cite{Schr09a}. In fact, each write operation is effectively an
erasure. (The quoted statistic is an average of $4,000$ correctable errors per
$128$ MB DIMM per year.) This gives a thermodynamic cost of $66~\kB T$, which
is markedly larger than Landauer's $\kB T\ln 2$ factor. It is also, just as
clearly, smaller by a factor of roughly $10^3$ than the contemporary energy
costs per logic operation displayed in Fig. \ref{fig:Stack}. These comparisons
are harbingers of the substantial effort that lies ahead to fully flesh-out and
fairly calibrate costs in Landauer's Stack.

\section{Erasure Thermodynamics}

Inequalities Eqs. (\ref{eq:TSP_work_bound}) and
(\ref{eq:TSP_work_bound_approx}) place severe constraints on the work required
to process information via time-symmetric control on memories. The question
remains, though, whether or not these bounds can actually be met by specific
protocols or if there might be still tighter bounds to be discovered.

To help answer this question, we turn to the case, originally highlighted by
Landauer \cite{Land61a}, of erasing a single bit of information. This
remarkably simple case of computing has held disproportionate sway in the
development of thermodynamic computing compared to other elementary operations.
The following does not deviate from this habit, showing, in fact, that there
remain fundamental issues. We explore this via two different implementations.
The first, described via two-state rate equations and the second with
an underdamped double-well potential---Landauer's original, preferred setting.

Suppose the SUS supports two (mesoscopic) memory states, labeled $\Left$ and
$\Right$. The task of a time-symmetric protocol that implements erasure is to
guide the SUS microscopic dynamics that starts with an initial $50-50$
distribution over the two memory states to a final distribution as biased as
possible onto the $\Left$ state. The logical function $\comp$ of perfect bit
erasure is attained when $\comp(\Left) = \comp(\Right) = \Left$, setting either
memory state to $\Left$. The probabilities of incorrectly sending an $\Left$
state to $\Right$ and an $\Right$ state to $\Right$ are denoted $\error_\Left$
and $\error_\Right$, respectively.

\renewcommand{\kbldelim}{(}\renewcommand{\kbrdelim}{)}

Error generation is described by the binary asymmetric channel
\cite{Cove06a}---the \emph{erasure channel} $\mathcal{E}$ with conditional probabilities:
\begin{align*}
\mathcal{E} = 
\kbordermatrix{
                       & \mathcal{M}_\tau = \Left & \mathcal{M}_\tau = \Right \\
\mathcal{M}_0 = \Left  & 1-\error_\Left        & \error_\Left \\
\mathcal{M}_0 = \Right & \error_\Right         & 1 - \error_\Right
}
  ~.
\end{align*}
For any erasure implementation, this Markov transition matrix gives the error
rate $\epsilon_\Left=\epsilon_{\Left \rightarrow \Right}$ from initial memory
state $\mathcal{M}_0=\Left$ and the error rate
$\epsilon_\Right=\epsilon_{\Right \rightarrow \Right}$ from the initial memory
state $\mathcal{M}_0=\Right$.

Noting first that $d(\mst, \mst) = 0$ generically, we then have:
\begin{align*}
d(\Left, \Right)
  & = \error_\Left \ln \frac{\error_\Left}{1-\error_\Right}
  ~, \\
  d(\Right, \Left)
  & = (1- \error_\Right) \ln \frac{1-\error_\Right}{\error_\Left}
  ~.
\end{align*}
So, the bound of Eq. (\ref{eq:TSP_work_bound}) simplifies to:
\begin{align}
	\beta \tspworkbound
	&= \frac{1}{2} \error_\Left \ln \frac{\error_\Left}{1 - \error_\Right}
		+ \frac{1}{2} (1 - \error_\Right) \ln \frac{1 -
		\error_\Right}{\error_\Left} \nonumber \\
	&= \Big( \frac{1}{2} - \erroravg \Big) \ln \frac{1-\error_\Right}{\error_\Left}
	\label{eq:erasure_work_bound}
	~,
\end{align}
where $\erroravg = (\error_\Left + \error_\Right)/2$ is the average error for
the process.

Notice further that $\comp(\comp(\Left)) = \Left$ but
$\comp(\comp(\Right)) \neq \Right$, indicating that only the computation on
$\Right$ is nonreciprocal. Therefore, the bound of Eq.
(\ref{eq:TSP_work_bound_approx}) simplifies to
\begin{align}
\beta \tspworkboundapprox
  & = \frac{1}{2} \ln(\error^{-1})
  ~.
\label{eq:erasure_work_bound_approx}
\end{align}

\subsection{Erasure with Two-state Rate Equations}
\label{sec:erasure two state}

A direct test of time-symmetric erasure requires only a simple two-state system
that evolves under a rate equation:
\begin{align}
& \frac{d \Pr(\MSt_t=m)}{dt}
\\& =\sum_{m'} \big[ r_{m' \rightarrow m}(t) \Pr(\MSt_t=m')-r_{m \rightarrow m'}(t) \Pr(\MSt_t=m) \big] \nonumber
  ,
\end{align}
obeying the Arrhenius equations:
\begin{align*}
r_{\Right \rightarrow \Left}(t) & = Ae^{-\Delta E_\Right(t)/ \kB T}
  ~\text{and} \\
r_{\Left \rightarrow \Right}(t) & = Ae^{-\Delta E_\Left(t) / \kB T}
  ~,
\end{align*}
where the states are labeled $\{\Left, \Right\}$ and the terms $\Delta
E_\Right(t)$ and $\Delta E_\Left(t)$ in the exponentials are the activation
energies to transit over the energy barrier at time $t$ for the Right and Left
wells, respectively.

These dynamics are a coarse-graining of thermal motion in a double-well
potential energy landscape $V(q,t)$ over the positional variable $q$ at time
$t$. Above, $A$ is an arbitrary constant, which is fixed for the dynamics.
$q^*_\Right$ and $q^*_\Left$ are the locations of the Right and Left potential
well minima, respectively. Thus, assuming that $q=0$ is the location of the
barrier's maximum between them, we see that the activation energies can be
expressed as $\Delta E_\Right(t) =V(0,t)-V(q^*_\Right,t)$ and $\Delta
E_\Left(t) =V(0,t)-V(q^*_\Left,t)$. By varying the potential energy extrema
$V(q^*_\Right,t)$, $V(q^*_\Left,t)$, and $V(0,t)$ we control the dynamics of
the observed variables $\{ \Left, \Right\}$ in much the same way as is done
with physical implementations of erasure where barrier height and tilt are
controlled in a double-well \cite{Jun14a}.

\begin{figure}[h]
\includegraphics[width=.9\columnwidth]{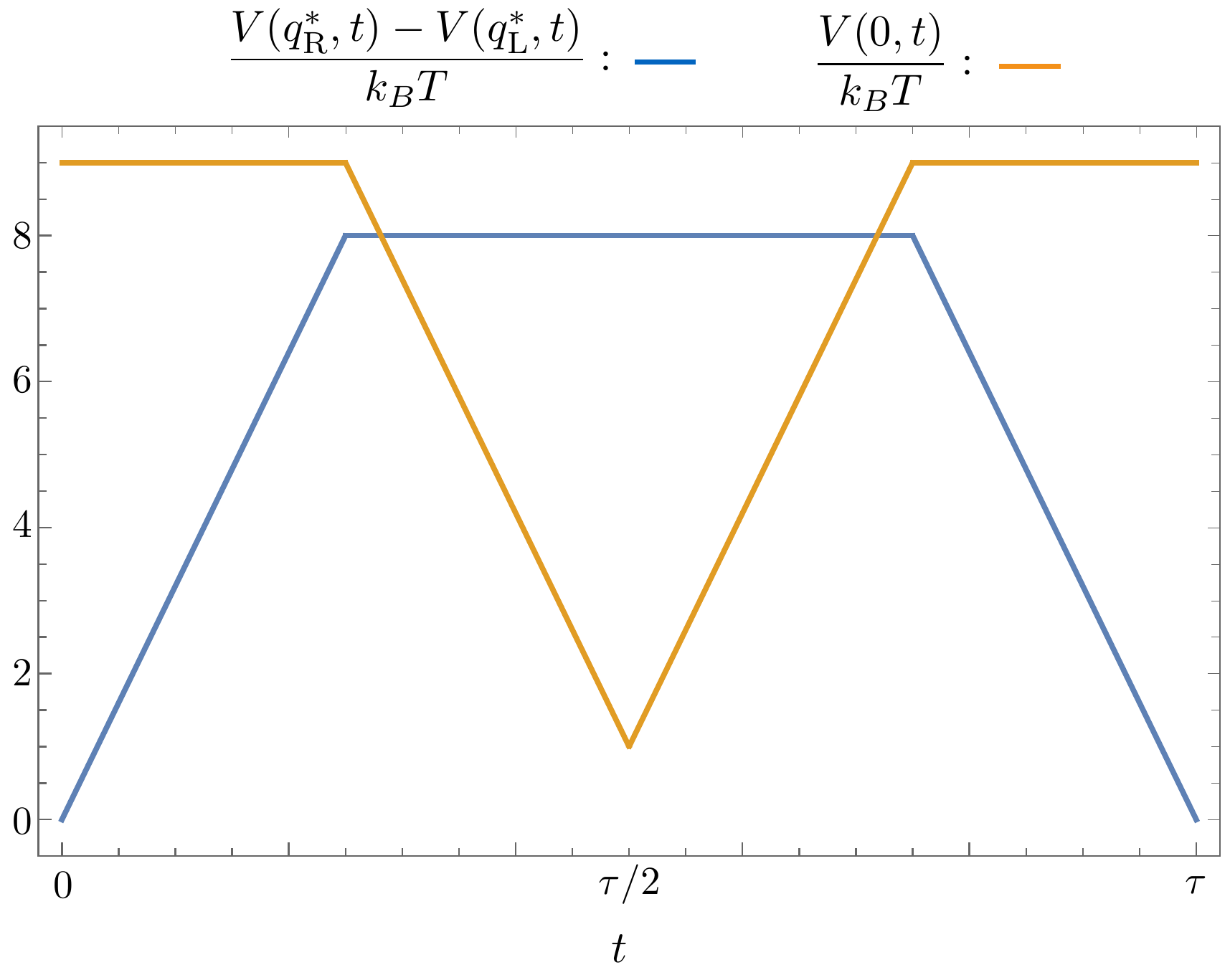}
\centering
\caption{Time-symmetric control protocol for implementing moderately-efficient
	erasure. This should be compared to Landauer's original time-symmetric
	protocol \cite{Land61a}. It starts by tilting---increasing the difference
	in potential energy $(V(q^*_\Right,t)-V(q^*_\Left,t))/ \kB T$ between
	$\Left$ and $\Right$. We increase this value such that transitions are more
	likely to go from $\Right$ to $\Left$. Then we reduce the barrier height
	$V(0,t)$ to increase the total flow rate.  Finally, we reverse the previous
	steps, cutting off the flow by raising the barrier, then untilting.
	}
\label{fig:ControlProtocol} 
\end{figure}

Deviating from previous investigations of efficient erasure, where Landauer's
bound was nearly achieved over long times \cite{Jun14a,Hong16a}, here the
constraint to symmetric driving over the interval $t \in (0, \tau)$
results in additional dissipated work. As Landauer described \cite{Land61a},
erasure can be implemented by turning on and off a tilt from $\Right$ to
$\Left$---a time symmetric protocol. However, to achieve higher accuracy, we
also lower the barrier while the system is tilted energetically towards the
$\Left$ well.

Consider a family of control protocols that fit the profile shown in Fig.
\ref{fig:ControlProtocol}. First, we increase the energy tilt from $\Right$ to
$\Left$ via the energy difference $V(q^*_\Right,t)-V(q^*_\Left,t)$ measured in
units of $\kB T$. This increases the relative probability of transitioning
$\Right$ to $\Left$. However, with the energy barrier at it's maximum height,
the transition takes quite some time. Thus, we reduce the energy barrier
$V(0,t)$ to its minimum height halfway through the protocol $t= \tau/2$. Then,
we reverse the protocol, raising the barrier back to its default height to hold
the probability distribution fixed in the well and untilt so that the system
resets to its default double-well potential.

\begin{figure}[h]
\includegraphics[width=\columnwidth]{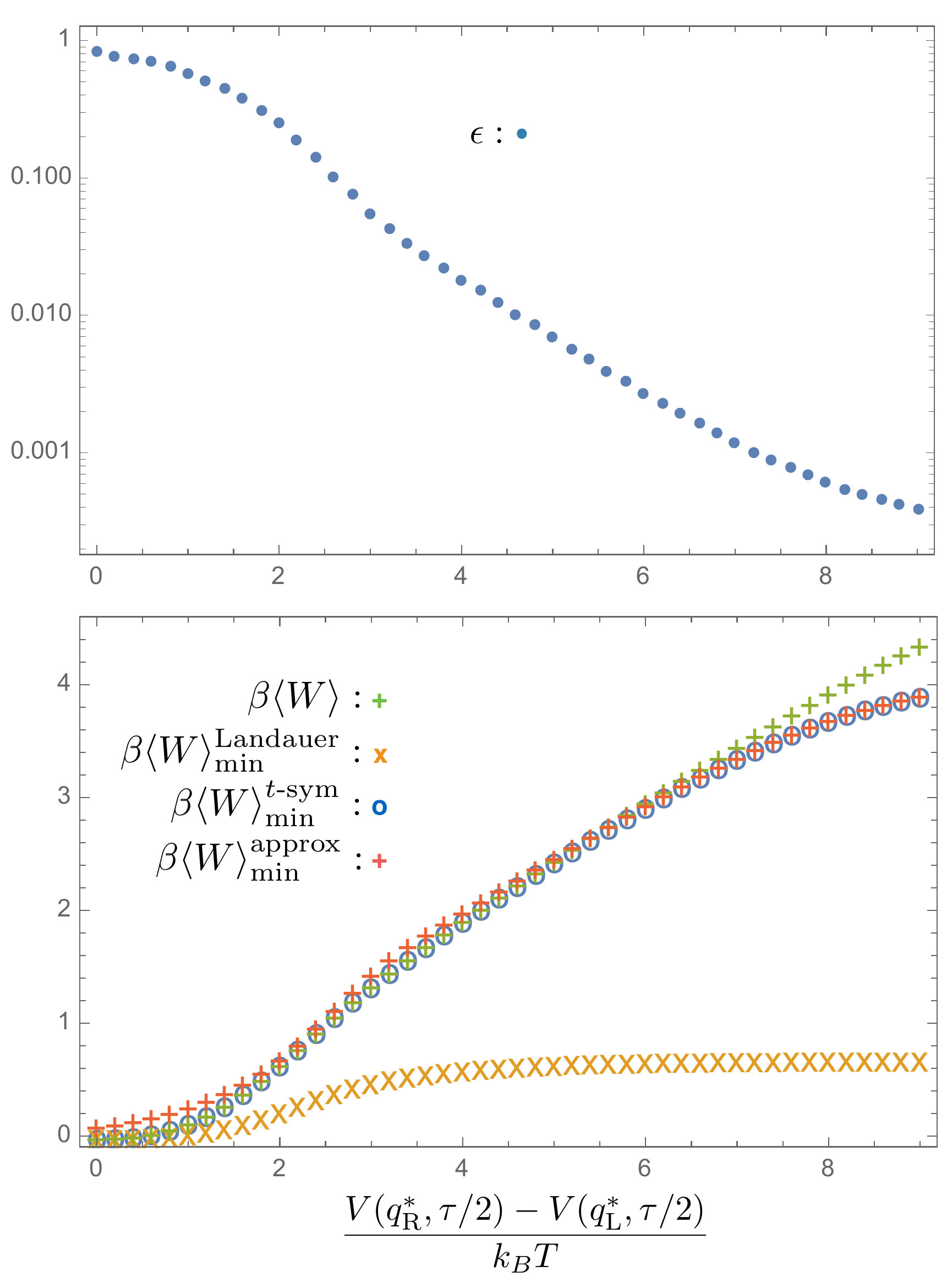}
\centering
\caption{(Top) Maximum error $\epsilon$ (blue dots) decreases approximately
	exponentially with increasing maximum tilt. The latter is given by the
	maximum energy difference between the right and left energy well
	$V(q^*_\Right,\tau/2)-V(q^*_\Left,\tau/2)$. (Bottom) Work $\langle W
	\rangle$ (green $+$s), scaled by the inverse temperature $\beta=1/ \kB T$,
	increases with increasing maximum tilt and decreasing error. The Landauer
	work bound $\braket{W}^\text{Landauer}_\text{min}$ (orange $\times$s) is a
	very weak bound, asymptoting to a constant value rather than continuing to
	increase, as the work does. The bound $\langle
	W\rangle^{t\text{-sym}}_\text{min}$ (blue circles) on time-symmetrically
	driven protocols, on the other hand, is a very tight bound for lower values
	of maximum tilt. The work deviates from the time-symmetric bound for higher
	tilts. Finally, the approximate bound $\braket{W}_\text{min}^\text{approx}$
	(red $+$s), which scales as $\ln \epsilon^{-1}$, is not an accurate bound
	over the entire range, but it very closely matches the exact time-symmetric
	bound $\langle W\rangle^{t\text{-sym}}_\text{min}$ for small $\epsilon$, as
	expected.
	}
\label{fig:WorkBounds}
\end{figure}

Increasing the maximum tilt---given by
$V(q^*_\Right,\tau/2)-V(q^*_\Left,\tau/2)$ at the halfway time---increases
erasure accuracy. Figure \ref{fig:WorkBounds} shows that the maximum error
$\epsilon = \max \{ \epsilon_\Right,\epsilon_\Left \}$ decreases nearly
exponentially with increased maximum energy difference between left and right,
going below $1$ error in every $1000$ trials for our parameter range. Note that
$\epsilon$ starts at a very high value (greater than $1/2$) for zero tilt,
since the probability $\epsilon_\Right=\epsilon$ of ending in the $\Right$ well
starting in the $\Right$ well is very high if there is no tilt to push the
system out of the $\Right$ well.

Figure \ref{fig:WorkBounds} also shows the relationship between the work and
the bounds described above. Given that our system consists of two states
$\{\Left, \Right \}$ and that we choose a control protocol that keeps the
energy on the left $V(q^*_\Left,t)$ fixed, the work (marked by green $+$s in
the figure) is \cite{Deff2013}:
\begin{align*}
\langle W \rangle & = \int_0^\tau dt \sum_s \Pr(\St_t=s)\partial_t V(s,t) \\
  & = \int_0^\tau dt \Pr(\MSt_t=\Right)\partial_tV(q^*_\Right,t)
  ~.
\end{align*}
This work increases almost linearly as the error reduces exponentially.

As a first comparison, note that the Landauer bound $\langle W
\rangle^\text{Landauer}_\text{min}=- \kB T\Delta H(\MSt_t)$ (marked by orange
$\times$s in the figure) is still valid. However, it is a very weak bound for
this time-symmetric protocol. The Landauer bound saturates at $\kB T \ln 2$.
Thus, the dissipated work---the gap between orange $\times$s and green
$+$s---grows approximately linearly with increasing tilt energy.

In contrast, Eq. (\ref{eq:erasure_work_bound})'s bound $\langle W
\rangle_\text{min}^{t\text{-sym}}$ for time symmetric protocols is much
tighter. The time symmetric bound is valid: marked by blue circles that all
fall below the calculated work (green $+$s). Not only is this bound much
stricter, but it almost exactly matches the calculated work for a large range of
parameters, with the work only diverging for higher tilts and lower error
rates.  

Finally, the approximate bound $\langle W \rangle_\text{min}^\text{approx} =
\frac{ \kB T}{2}\ln \epsilon^{-1}$ (marked by red $+$s) of Eq.
(\ref{eq:erasure_work_bound_approx}), which captures the error scaling, behaves
as expected. The error-dependent work bound nearly exactly matches the exact
bound for low error rates on the right side of the plot and effectively bounds
the work. For lower tilts, this quantity does not bound the work and is not a
good estimate of the true bound, but this is consistent with expectations for
high error rates. This approximation should only be employed for very reliable
computations, for which it appears to be an excellent estimate. Thus, the
two-level model of erasure demonstrates that the time-symmetric control bounds
on work and dissipation are reasonable in both their exact and approximate
forms at low error rates.

\subsection{Erasure with an Underdamped Double-well Potential}
\label{sec:erasure langevin}

The physics in the rate equations above represents a simple model of a bistable
thermodynamic system, which can serve as an approximation for many different
bistable systems. One possible interpretation is a coarse-graining of the
Langevin dynamics of a particle moving in a double-well potential. To explore
the broader validity of the error--dissipation tradeoff, here we simulate the
dynamics of a stochastic particle coupled to a thermal environment at constant
temperature and a work reservoir via such a 1D potential. Again, we find that
the time-symmetric bounds are much tighter than Landauer's, reflecting the
error--dissipation tradeoff of this control protocol class.

Consider a one-dimensional particle with position and momentum in an external
potential and in thermal contact with the environment at temperature $T$.
We consider a protocol architecture similar to that of Sec.
\ref{sec:erasure two state}, but with additional passive substages at the
beginning middle and end: (i) hold the potential in the symmetric
double-well form, (ii) positively tilt the potential, (iii) completely drop the
potential barrier between the two wells, (iv) hold the potential while it is
tilted with no barrier, (v) restore the original barrier, (vi) remove the
positive tilt, restoring the original symmetric double-well, and (vii) hold the
potential in this original form.

\begin{figure}[h]
\includegraphics[width=\columnwidth]{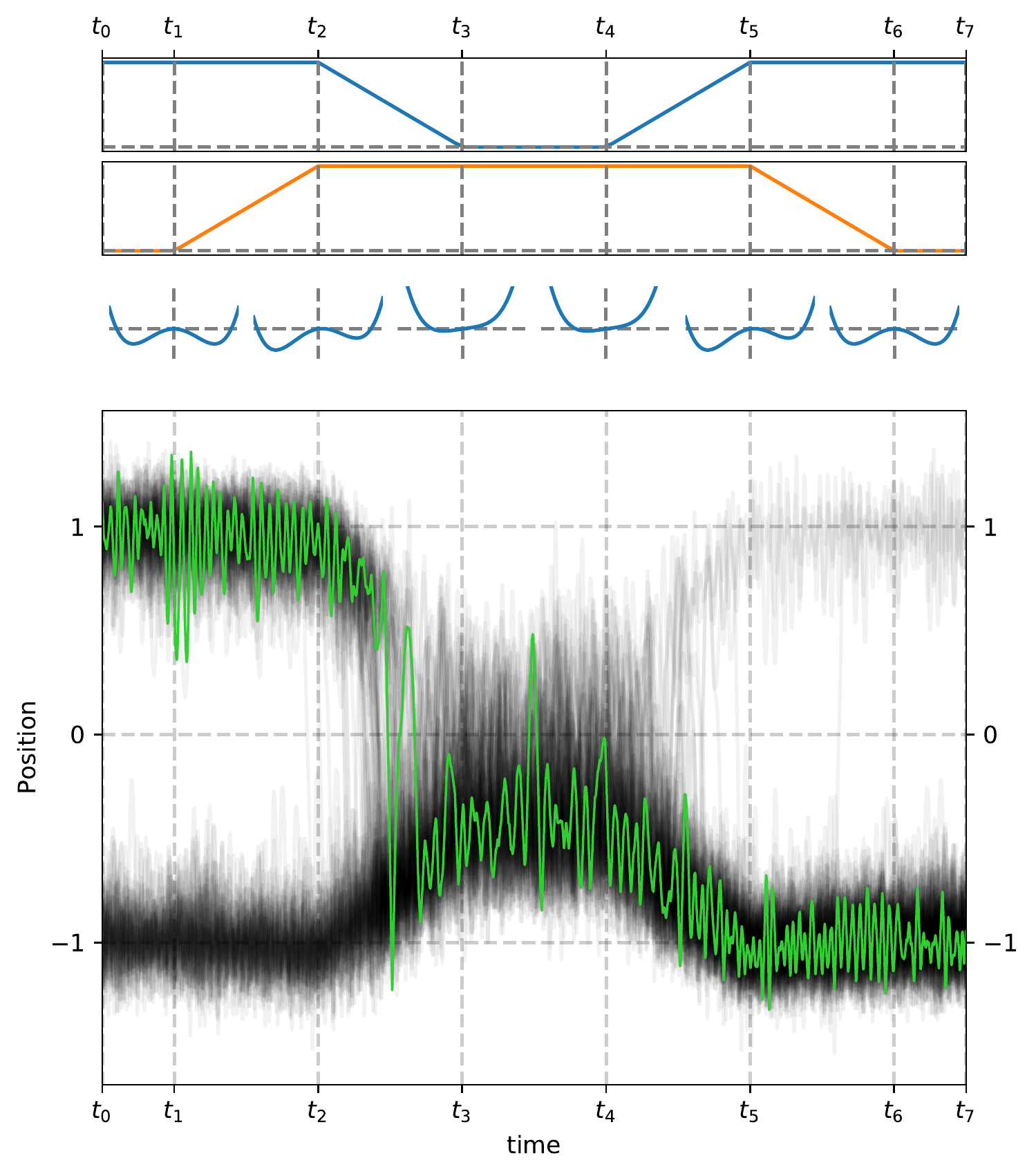}
\caption{Erasure via an underdamped double-well potential: Protocol functions
	$b(t)$ (top panel, blue) and $c(t)$ (second panel, orange) are symmetric in
	time, guaranteeing the potential function (third panel) to evolve
	symmetrically in time. Due to the spatial asymmetry in the potential over
	the majority of the protocol, however, erasure to state $\Left$ ($x<0$)
	typically occurs, evidenced by the evolution of the system position for
	$100$ randomly-chosen trajectories (bottom panel, black). The $\Left$ and
	$\Right$ states merge into one between times $t_2$ and $t_3$ and separate
	again between times $t_4$ and $t_5$. A single trajectory (bottom panel,
	green) shows the typical behavior of falling into the $x<0$ region by time
	$t_3$ and remaining there when the $\Left$ state is reintroduced for the
	rest of the protocol.
	}
\label{fig:langevin_sim} 
\end{figure}

As a function of position $q$ and time $t$, the potential then takes the form:
\begin{align*}
V(q, t) = a q^4 - b_0 b_f (t) q^2 + c_0 c_f(t) q
 ~,
\end{align*}
with constants $a, b_0, c_0 > 0$. The protocol functions $b_f(t)$ and $c_f(t)$
evolve in a piecewise linear, cyclic, and time-symmetric manner according to
Table \ref{tab:langevin_protocol}, where $t_0, t_1, \ldots , t_7 = 0, \tau/12,
3\tau/12, 5\tau/12, 7\tau/12, 9\tau/12, 11\tau/12, \tau$. The potential thus
begins and ends in a symmetric double-well configuration with each well
defining a memory state. During the protocol, though, the number of metastable
regions is temporarily reduced to one. Figure \ref{fig:langevin_sim} (top three
panels) shows the protocol functions over time as well as the resultant
potential function at key times for one such set of protocol parameters; see
nondimensionalization in App. \ref{app:Langevin Simulations of Erasure}. At
any time, we label the metastable regions from most negative position to most
positive the $\Left$ state and, if it exists, the $\Right$ state.

\begin{table}[ht]
\begin{center}
{\setlength{\extrarowheight}{3pt}\begin{tabular}{|c|| c @{\hspace{-1pt}} c @{\hspace{-1pt}} c @{\hspace{-1pt}} c
@{\hspace{-1pt}}  c @{\hspace{-1pt}} c @{\hspace{-1pt}} c @{\hspace{-1pt}} c
@{\hspace{-1pt}} c @{\hspace{-1pt}} c @{\hspace{-1pt}} c @{\hspace{-1pt}} c
@{\hspace{-1pt}} c @{\hspace{-1pt}} c @{\hspace{-1pt}} c|}
\hline
$t$  & $t_0$ && $t_1$ && $t_2$ && $t_3$ && $t_4$ && $t_5$ && $t_6$ && $t_7$ \\
\hline
$b_f(t)$ &$\biggr\vert$& 1 &$\biggr\vert$& 1 &$\biggr\vert$&
$\frac{t_3-t}{t_3-t_2}$ &$\biggr\vert$& 0 &$\biggr\vert$& $\frac{t-t_4}{t_5-t_4}$
&$\biggr\vert$& $1$ &$\biggr\vert$& $1$ &$\biggr\vert$ \\
$c_f(t)$ & $\biggr\vert$& 0 &$\biggr\vert$& $\frac{t-t_1}{t_2-t_1}$ &
$\biggr\vert$& 1&$\biggr\vert$& $1$ &$\biggr\vert$& $1$ &$\biggr\vert$
& $\frac{t_6 - t}{t_6 - t_5}$ &$\biggr\vert$ &$0$ &$\biggr\vert$ \\
\hline
    \end{tabular}}
  \end{center}
\caption{Erasure protocol.
  }
\label{tab:langevin_protocol}
\end{table}

We simulate the motion of the particle with underdamped Langevin dynamics:
\begin{align*}
dq & = v dt \\
m dv & = - \left( \frac{\partial}{\partial q} V(q, t) + \lambda v \right) dt
  + \sqrt{2 \kB T\lambda}\, r(t) \sqrt{dt}
  ~,
\end{align*}
where $\kB$ is Boltzmann's constant, $\lambda$ is the coupling between the
thermal environment and particle, $m$ is the particle's mass, and $r(t)$ is a
memoryless Gaussian random variable with $\langle r(t) \rangle = 0$ and
$\langle r(t) r(t') \rangle = \delta(t-t')$. The particle is initialized to be
in global equilibrium over the initial potential $V(\cdot, 0)$. Figure
\ref{fig:langevin_sim} (bottom panel) shows $100$ randomly-chosen resultant
trajectories for a choice of process parameters.

\begin{figure}[h]
\includegraphics[width=\columnwidth]{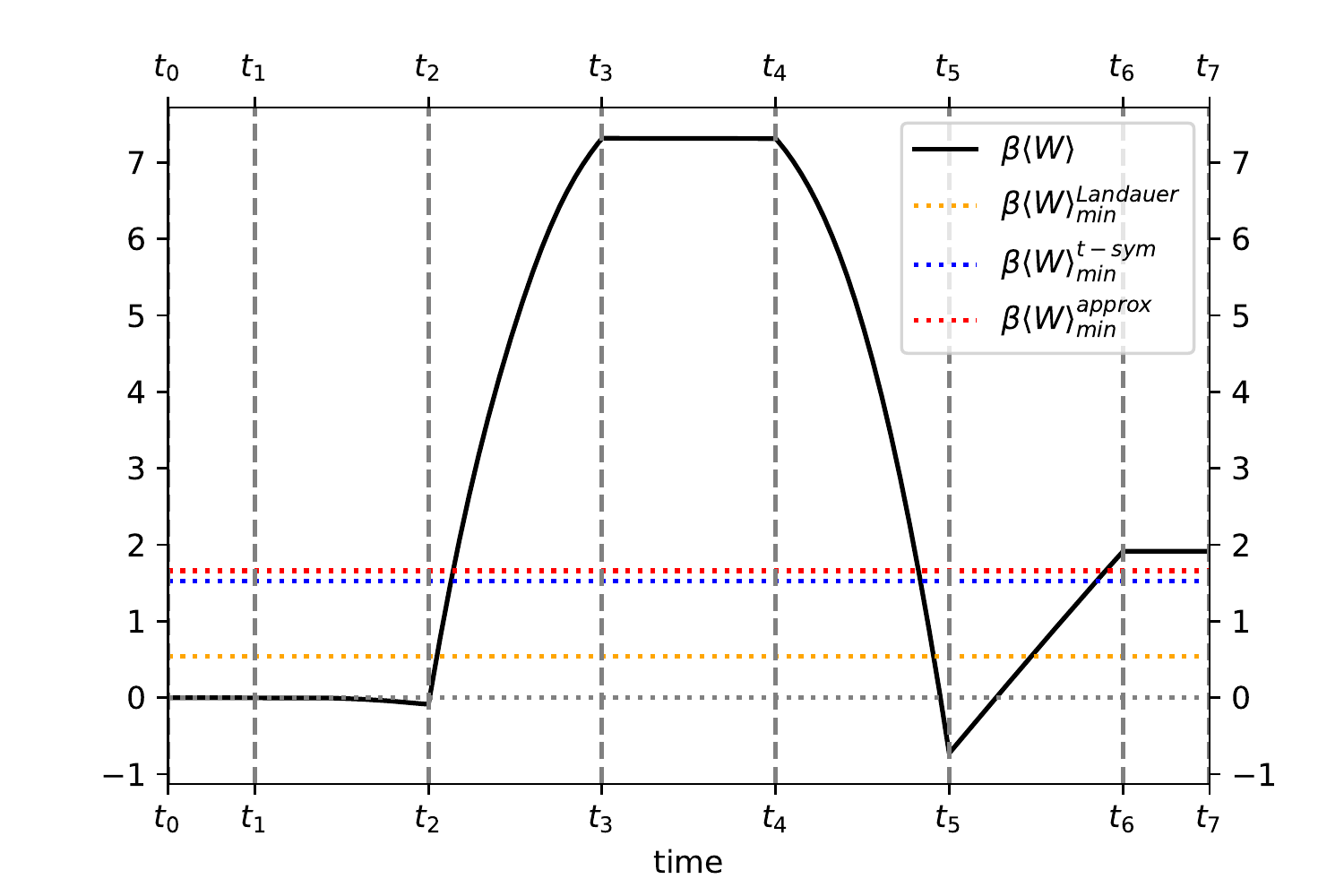}
\caption{Average work in $\kB T$ over time for an erasure (black). Calculated
	from the simulation-estimated values $\error_L$ and $\error_R$, Landauer's
	bound is given by the dashed yellow line and our approximate and exact
	bounds (Eqs. (\ref{eq:erasure_work_bound_approx}) and
	(\ref{eq:erasure_work_bound})) are given in dashed red and blue lines,
	respectively.
	}
\label{fig:langevin_work} 
\end{figure}

The work done on a single particle over the course of the protocol with
trajectory $\{q(t)\}_t$ is \cite{Deff2013}:
\begin{align*}
W = \int_0^\tau dt \frac{\partial V(q, t)}{\partial t} \Big|_{q=q(t)}
 ~.
\end{align*}
Figure \ref{fig:langevin_work} shows the net average work over time for an
erasure process, comparing it to (i) the Landauer bound, (ii) the exact bound
of Eq. (\ref{eq:erasure_work_bound}), and (iii) the approximate bound of Eq.
(\ref{eq:erasure_work_bound_approx}). Notice that the final net average work
lies above all three, as it should and that the time-symmetric bounds presented
here are tighter than Landauer's.

We repeat this comparison for an array of different parameters for the erasure
protocol.  As described in App. \ref{app:Langevin Simulations of Erasure}, we
vary features of the dynamics---including mass $m$, temperature $T$, coupling
to the heat bath $\lambda$, duration of control $\tau$, maximum depth of the
potential energy wells, and maximum tilt between the wells.
Nondimensionalization reduces the relevant parameters to just four, allowing us
to explore a broad swathe of possible physical erasures with 735 different
protocols. For each protocol, we simulate 100,000 trajectories to estimate the
work cost and errors $\epsilon_\text{R}$ and $\epsilon_\text{L}$ of the
operation.

\begin{figure}[h]
\includegraphics[width=\columnwidth]{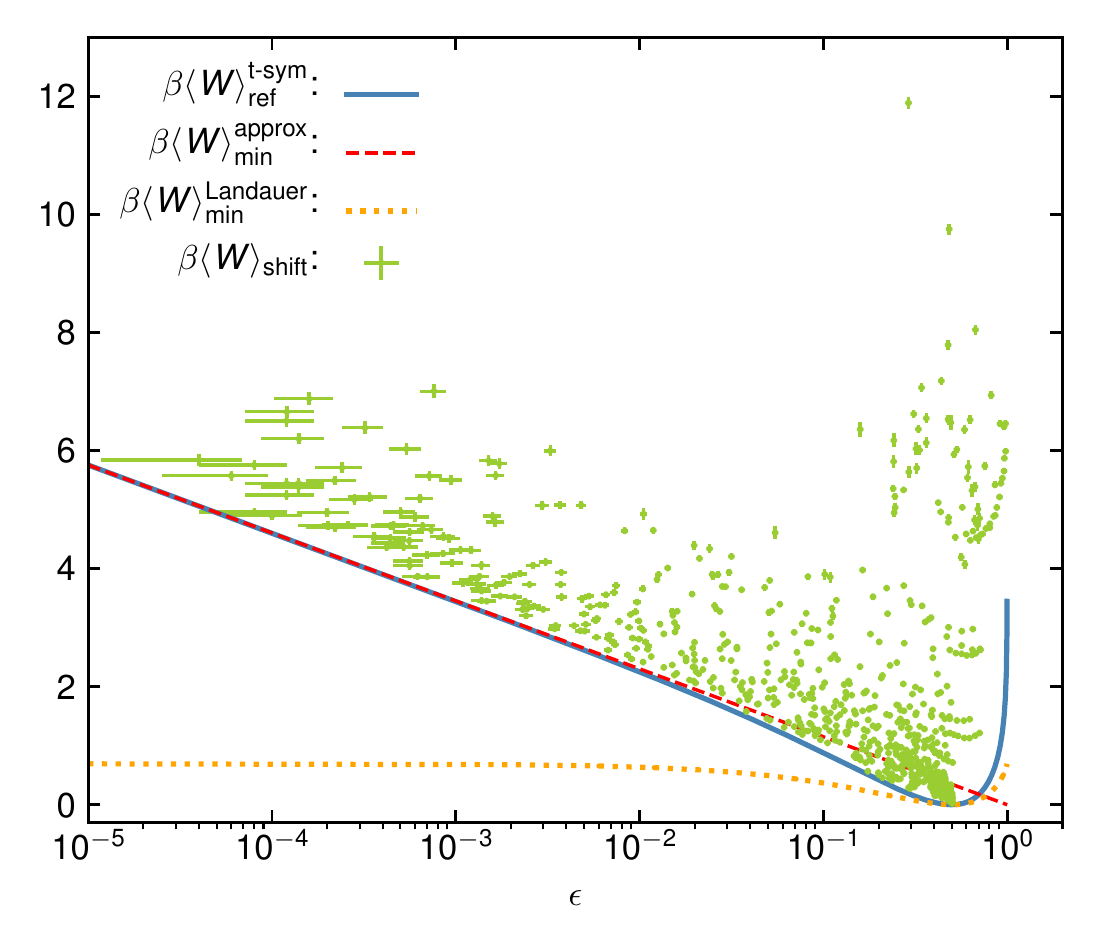}
\caption{Reference bound $\langle W \rangle_\text{ref}^{t\text{-sym}}$
	(blue line) lower bounds all of the shifted works $\langle W
	\rangle_\text{shift}$ (green markers), often quite tightly. The approximate
	bound $\langle W \rangle_\text{min}^{\text{approx}}$ (red dashed line)
	rapidly converges with decreasing error to $\langle W
	\rangle_\text{ref}^{t\text{-sym}}$. Time-asymmetric protocols can do
	better, needing only to satisfy Landauer's bound $\langle W
	\rangle_\text{min}^\text{Landauer}$ (orange dotted line).
	}
\label{fig:ErrorAndDissipationInErasure} 
\end{figure}

Figure~\ref{fig:ErrorAndDissipationInErasure} compares the work spent for each
of the 735 erasure protocols to the sampled maximum error $\epsilon =
\max(\epsilon_\Left,\epsilon_\Right)$. Each protocol corresponds to a green
cross, whose vertical position corresponds to the shifted work $\langle
W\rangle_\text{shift}$, which accounts for inhomogeneities in the error rate.
Note that the exact bound $\langle W \rangle_\text{min}^{t\text{-sym}}$ from
Eq.~(\ref{eq:erasure_work_bound}) reduces to a simple relationship between work
and error tolerance $\epsilon$ when the errors are homogeneous
$\epsilon_\Right=\epsilon_\Left=\epsilon$:
\begin{align*}
\langle W\rangle ^{t\text{-sym}}_\text{ref}=\left(\frac{1}{2}-\epsilon \right) \ln \frac{1-\epsilon}{\epsilon}
  ~,
\end{align*}
which we plot with the blue curve in
Fig.~\ref{fig:ErrorAndDissipationInErasure}. The cost of inhomogeneities in
the error is evaluated by the difference between this reference bound and the
exact work bound. This cost is added to the calculated work for each protocol
to determine the shifted work:
\begin{align*}
\langle W\rangle_\text{shift}=\langle W \rangle +\langle W\rangle ^{t\text{-sym}}_\text{ref}-\langle W\rangle ^{t\text{-sym}}_\text{min}
  ~,
\end{align*}
such that the vertical distance between $\langle W \rangle_\text{shift}$ and
$\langle W \rangle_\text{ref}^{t\text{-sym}}$ in
Fig.~\ref{fig:ErrorAndDissipationInErasure} gives the true difference
$\langle W \rangle - \langle W \rangle_\text{min}^{t\text{-sym}}$ between the
average sampled work and exact bound for the simulated protocol.

Figure~\ref{fig:ErrorAndDissipationInErasure} shows that the shifted average
works for all of the simulated protocols in green, including error bars, all
lay above the reference work bound in blue. Thus, we see that all simulated
protocols satisfy the bound $\langle W \rangle \geq \langle W
\rangle_\text{min}^{t\text{-sym}}$.
Furthermore, many simulated protocols end up quite close to their exact bound.
There are protocols with small errors, but they have larger average works. The
error--dissipation tradeoff is clear.

The error--dissipation tradeoff is further illustrated in
Fig.~\ref{fig:ErrorAndDissipationInErasure} by the red line, which describes
the low-$\epsilon$ asymptotic bound $\langle W
\rangle_\text{min}^\text{approx}$ given by
Eq.~(\ref{eq:erasure_work_bound_approx}). In this semi-log plot, it rather
quickly becomes an accurate approximation for small error.

Finally, Fig.~\ref{fig:ErrorAndDissipationInErasure} plots the Landauer bound
$\langle W \rangle_\text{min}^\text{Landauer}$ as a dotted orange line. It is
calculated using the final probability of the $\Right$ mesostate. The bound is
weaker than that set by $\langle W \rangle_\text{ref}^{t\text{-sym}}$. As
$\epsilon \to 0$, the gap between $\langle W \rangle_\text{ref}^{t\text{-sym}}$
and $\langle W \rangle_\text{min}^\text{Landauer}$ in
Fig.~\ref{fig:ErrorAndDissipationInErasure} relentlessly increases. The stark
difference in the energy scale of the time-symmetric bounds developed here and
that of the looser Landauer bound shows a marked tightening of thermodynamic
bounds on computation.

Notably, the protocol Landauer originally proposed to erase a bit requires
\emph{significantly more work} than his bound $\kB T \ln 2$ to reliably erase a
bit. This extra cost is a direct consequence of his protocol's time symmetry.
It turns out that time-\emph{asymmetric} protocols for bit erasure have been
used in experiments that more nearly approach Landauer's bound~\cite{Dill09,
Jun14}. Although, it is not clear to what extent time asymmetry was an
intentional design constraint in their construction, since there was no general
theoretical guidance until now for why time-symmetry or asymmetry should
matter. Figures~\ref{fig:ErrorAndDissipationInErasure} and \ref{fig:WorkBounds}
confirm that Ref. \cite{Jun14}'s time-asymmetric protocol for bit
erasure---where the barrier is lowered before the tilt, but then raised before
untilting---is capable of reliable erasure that is more thermodynamically
efficient than any time-symmetric protocol could ever be.

These underdamped simulations drive home the point that our bounds are
independent of the details of the dynamics used for computation. Our results
are very general in that regard. As long as the system \emph{starts} metastable
and is then driven by a time-symmetric protocol, the error--dissipation
tradeoff quantifies the minimal dissipation that will be incurred (for a
desired level of computational accuracy) by the time the system relaxes again
to metastability.

\section{Conclusion}

We adapted Ref. \cite{Riec19b}'s thermodynamic analysis of time-symmetric protocols to give a detailed analysis of the trade-offs between accuracy and dissipation encountered in erasing information.

Reference \cite{Riec19b} showed that time symmetry and metastability together
imply a generic error--dissipation tradeoff. The minimal work expected for a
computation $\mathcal{C}$ is the average nonreciprocity. In the low-error
limit---where the probability of error must be much less than unity ($\epsilon
\ll 1$)---the minimum work diverges according to: 
\begin{align*}
\beta \langle W \rangle_\text{min}^\text{approx} 
= \bigl\langle \IverL \mathcal{C}(\mathcal{C}(\MSt_0) )
  \neq \MSt_0  \IverR \bigr\rangle_{\MSt_0} \ln(\epsilon^{-1} )
\end{align*}
Of all of this work, only the meager Landauer cost $\Delta \Shannon(\MSt_t)$,
which saturates to some finite value as $\epsilon \to 0$, can be
thermodynamically recovered in principle. Thus, irretrievable dissipation
scales as $\ln(\epsilon^{-1} )$. The reciprocity coefficient $ \bigl\langle
\IverL \mathcal{C}(\mathcal{C}(\MSt_0))\neq \MSt_0  \IverR
\bigr\rangle_{\MSt_0}$ depends only on the deterministic computation to be
approximated. This points out likely energetic inefficiencies in current
instantiations of reliable computation. It also suggests that time-asymmetric
control may allow more efficient computation---but only when time-asymmetry is
a free resource, in contrast to modern computer architecture.

The results here verified these general conclusions for erasure, showing in
detail how tight the bounds can be and, for high-reliability thermodynamic
computing, how they overwhelm Landauer's. Despite the almost
universal focus on information erasure as a proxy for all of computing, we now
see that there is a wide diversity of costs in thermodynamic computing. Looking
to the future, these costs must be explored in detail if we are to design and
build more capable and energy efficient computing devices. Beyond
engineering and sustainability concerns, explicating Landauer's Stack will go a
long way to understanding the fundamental physics of computation---one of
Landauer's primary goals \cite{Land81a}. In this way, we now better appreciate
the suite of thermodynamic costs---what we called Landauer's Stack---that
underlies modern computing. 

\section*{Acknowledgments}
\label{sec:acknowledgments}

The authors thank the Telluride Science Research Center for hospitality during
visits and the participants of the Information Engines Workshops there for
helpful discussions. JPC acknowledges the kind hospitality of the Santa Fe
Institute, Institute for Advanced Study at the University of Amsterdam, and
California Institute of Technology for their hospitality during visits. This
material is based upon work supported by, or in part by, FQXi Grant number
FQXi-RFP-IPW-1902, Templeton World Charity Foundation Power of Information
fellowship TWCF0337, and U.S. Army Research Laboratory and the U.S. Army
Research Office under contracts W911NF-13-1-0390 and W911NF-18-1-0028.

\appendix

\section{Langevin Simulations of Erasure}
\label{app:Langevin Simulations of Erasure}

To help simulate a wide variety of protocols, we first nondimensionalize the
equations of motion, using variables:
\begin{align*}
q' &= \sqrt{\frac{2a}{b_0}} q ~, \;
t' = \frac{2a \kB T}{b_0 \lambda} t ~, \;
v' = \frac{\lambda}{\kB T}\sqrt{\frac{b_0}{2a}} v ~,~\text{and} \\
 V' &= \frac{1}{\kB T} V
 ~.
\end{align*}
Note that the position scale $\sqrt{b_0 / 2a}$ is the distance from
zero to either well minima in the default potential $V(\cdot, 0) = V(\cdot,
\tau)$. Substitution then provides the following nondimensional equations:
\begin{align*}
dq' &= v' dt' \\
m' dv' &= - \left(\frac{\partial}{\partial q'} V'(q', t') + v' \right) dt'
  + \sqrt{2} r(t') \sqrt{dt'}
  ~,
\end{align*}
with:
\begin{align*}
m' &= \frac{2 a m \kB T}{b_0 \lambda^2}
 ~,
\end{align*}
which is the first nondimensional parameter to specify an erasure process.

The nondimensional potential can be expressed as:
\begin{align*}
V'(q', t') &= \alpha \Big( q'^4 - 2 b_f'(t') q'^2 + \zeta c_f'(t') q' \Big)
 ~,
\end{align*}
where:
\begin{align*}
\alpha = \frac{b_0^2}{4a \kB T}  \; ~\text{and}~ \;
\zeta = c_0 \sqrt{\frac{2a}{b_0^3}}
\end{align*}
are two more nondimensional parameters to specify and:
\begin{align*}
b_f'(t') = b_f \left( \frac{2a \kB T}{b_0 \lambda} t \right) \; ~\text{and}~ \;
c_f'(t') = c_f \left( \frac{2a \kB T}{b_0 \lambda} t \right)
\end{align*}
simply express $b_f$ and $c_f$ with the nondimensional time as input.
The fourth and final nondimensional parameter is the nondimensional total time:
\begin{align*}
\tau' = \frac{2a \kB T}{b_0 \lambda} \tau
 ~.
\end{align*}

To explore the space of possible underdamped erasure dynamics, we simulate
$735$ different protocols, determined by all combinations of the following
values for the four nondimensional parameters: $m' \in \{ 0.25, 1.0, 4.0 \}$,
$\alpha \in \{ 2, 4, 7, 10, 12 \}$, $\zeta \in \{ 0.1, 0.2, 0.3, 0.4, 0.5, 0.6,
0.7 \}$, and $\tau' \in \{ 4, 8, 16, 32, 64, 128, 256 \}$. $100,000$ trials of
each parameter set were simulated.  For the simulations of Figs.
\ref{fig:langevin_sim} and \ref{fig:langevin_work}, we set $m'=1$, $\alpha=7$,
$\zeta=0.4$, and $\tau'=100$. Figure~\ref{fig:ErrorAndDissipationInErasure}
shows that the  (error, work) pairs obtained for these various dynamics fill
in the region allowed by our time-symmetric bounds. These bounds can indeed be
tight, but it is quite possible to waste more energy if the computation is not
tuned for energetic efficiency.

To update particle position and velocity each time step, we used the
fourth-order Runge-Kutta integration for the deterministic portion of the
equations of motion and a simple Euler method in combination with a Gaussian
number generator for the stochastic portion. To determine the time step size,
we considered a range of possible time steps for $81$ of the possible $735$
parameter sets and looked for convergence of the sampled average works and
maximum errors $\epsilon$, again using $100,000$ trials per parameter set.

The maximum errors were stable over the whole range of tested step sizes.
Looking with decreasing step size, the final step size of $0.0025$ was chosen
when the average works stopped fluctuating within $5 \sigma$ of their
statistical errors for all $81$ parameter sets. The error bars presented for
the average works in Fig.~\ref{fig:ErrorAndDissipationInErasure} were then
generously set to be $5$ times the estimated statistical errors, which were
each obtained by dividing the sampled standard deviation by the square root of
the number of trials. Error bars for the maximum errors were set to be the
statistical errors of $\epsilon_\text{L}$ or $\epsilon_\text{R}$, depending on
which was the maximum, whose statistical errors were obtained by assuming
binomial statistics.


\end{document}